\documentclass[aps,prd,draft,amsmath,amssymb,preprintnumbers,nofootinbib,a4paper,11pt]{revtex4-1}
\pdfoutput=1
\usepackage{amsthm}
\usepackage{graphicx}
\usepackage{color}
\newcommand{\beq}{\begin{eqnarray}}
\newcommand{\eeq}{\end{eqnarray}}

\newcommand{\bmp}{\noindent\begin{minipage}{16cm}}
\newcommand{\emp}{\end{minipage}\vskip 7mm} 

\usepackage{dcolumn}
\usepackage{bm}
\usepackage{bbm}
\usepackage{subfigure}
\usepackage{pxfonts}
\usepackage[usenames,dvipsnames]{xcolor}

\theoremstyle{definition}

\theoremstyle{plain}

\usepackage{epsfig}

\usepackage[margin=5pt, font=normalsize,labelfont=bf,justification=raggedright]{caption}
\usepackage{youngtab}
\usepackage{slashed}

\definecolor{rossoCP3}{cmyk}{0,.88,.77,.40}

\def\lsim{\mathrel{\rlap{\lower4pt\hbox{\hskip1pt$\sim$}}
    \raise1pt\hbox{$<$}}}                
\def\gsim{\mathrel{\rlap{\lower4pt\hbox{\hskip1pt$\sim$}}
    \raise1pt\hbox{$>$}}}                

\newcommand{\drawsquare}[2]{\hbox{%
\rule{#2pt}{#1pt}\hskip-#2pt
\rule{#1pt}{#2pt}\hskip-#1pt
\rule[#1pt]{#1pt}{#2pt}}\rule[#1pt]{#2pt}{#2pt}\hskip-#2pt
\rule{#2pt}{#1pt}}

\newcommand{\Yfund}{\raisebox{-.5pt}{\drawsquare{6.5}{0.4}}}
\newcommand{\Ysymm}{\Yfund\hskip-0.4pt%
                    \Yfund}
\def\symm{\Ysymm}
\def\bsymm{\overline{\Ysymm}}

\def\drawbox#1#2{\hrule height#2pt
        \hbox{\vrule width#2pt height#1pt \kern#1pt
              \vrule width#2pt}
              \hrule height#2pt}

\def\Asym#1#2{\vcenter{\vbox{\drawbox{#1}{#2}
              \kern-#2pt       
              \drawbox{#1}{#2}}}}

\def\asymm{\Asym{6.4}{0.3}}

\def\basymm{\overline{\asymm}}

  \newcommand{\be}{\begin{eqnarray}}
\newcommand{\ee}{\end{eqnarray}}

\begin{document}
\title{\LARGE \color{rossoCP3} Composite Inflation \\ from \\ Super Yang-Mills, Orientifold and One-Flavor QCD }

\author{{\sc {Phongpichit Channuie}}$^{\color{rossoCP3}{\varheartsuit} {\clubsuit} }$}\email{channuie@cp3.dias.sdu.dk} 
\author{\sc {Jakob Jark J\o rgensen}$^{\color{rossoCP3}{\varheartsuit}}$}\email{joergensen@cp3.dias.sdu.dk} 
\author{\sc  {Francesco Sannino}$^{\color{rossoCP3}{\varheartsuit}}$}\email{sannino@cp3.dias.sdu.dk} 
 \affiliation{
$^{\color{rossoCP3}{\varheartsuit}} ${ \color{rossoCP3}  \rm CP}$^{\color{rossoCP3} \bf 3}${\color{rossoCP3}\rm-Origins} \& the Danish Institute for Advanced Study {\color{rossoCP3} \rm DIAS},\\ 
University of Southern Denmark, Campusvej 55, DK-5230 Odense M, Denmark.
}
  \affiliation{
$^{\color{rossoCP3}{\clubsuit}} $Dept. of Physics, Faculty of Liberal Arts and Science,
Kasetsart University, Kamphaeng Saen campus, Nakhon Pathom, 73140, Thailand.}
\begin{abstract}
Recent investigations have shown that inflation can be driven by four-dimensional strongly interacting theories non-minimally coupled to gravity. We explore this paradigm further by considering composite inflation driven by orientifold field theories. The advantage of using these theories resides in the fact that at large number of colors they feature certain super Yang-Mills properties. In particular we can use for inflation the bosonic part of the Veneziano-Yankielowicz effective theory. Furthermore, we include the 1/N as well as fermion mass corrections at the effective Lagrangian level allowing us to explore the effects of these corrections on the inflationary slow-roll parameters. Additionally the orientifold field theory with fermionic matter transforming according to the two-index antisymmetric representation for three colors is QCD.  Therefore this model can be interpreted as a new non-minimally coupled QCD theory of inflation. The scale of composite inflation, for all the models presented here, is of the order of $10^{16}$~GeV. Unitarity studies of the inflaton scattering suggest that the cutoff of the model is at the Planck scale.  \\ 
[.1cm]
{\footnotesize  \it Preprint: CP$^3$-Origins-2012-25 \& DIAS-2012-26}
 \end{abstract}

\maketitle

\section{Introduction} 
 
 Little is known about the mechanism underlying the inflationary physics  \cite{
 Guth:1980zm,Linde:1981mu,Albrecht:1982wi} postulated to occur soon after the birth of our Universe. The simplest models of inflation make use of  elementary scalar fields. However, the fundamental constituents of space-time are spinors. Scalars can be built out of the fundamental spinors but not vice-versa. It is therefore interesting to investigate whether the inflaton field can emerge as a composite state of a new strongly interacting gauge theory \cite{Channuie:2011rq,Bezrukov:2011mv}. Holographic models of composite inflation are also being currently investigated \cite{Evans:2010tf,Evans:2012jx}. 
  
To investigate this class of inflationary models one uses low energy effective theories constrained using the global symmetries of the theory as well as conformal symmetries. There is an interesting class of models featuring only fermionic and gluonic degrees of freedom for which one can use also supersymmetric relations \cite{Armoni:2003gp,Armoni:2003fb} to further constrain the low energy effective theory. These are gauge theories with fermionic matter transforming according to the two-index representation of the underlying  $SU(N)$ gauge dynamics.  These theories can be connected to $N=1$ super Yang-Mills (SYM) at large number of colors and are also known as orientifold theories\footnote{For matter transforming according to the two-index antisymmetric representation it was recognized some time ago \cite{Corrigan:1979xf} that these theories can be viewed as a large N generalization of QCD different from the 't Hooft large $N$.  Yet another distinct large $N$ generalization of QCD was introduced in \cite{Ryttov:2005na}}. The field content of the theories is reported in the Table \ref{table}.
\begin{table}[h]
\begin{center}
\begin{minipage}{2.8in}
\begin{tabular}{c||ccc }
 & $SU(N)$ & $U_V(1)$ & $U_A(1)$  \\
  \hline \hline 
${\psi_{\{ij\}}}$& $\symm$ & $1$ & $1$  \\
 $\widetilde{\psi}^{\{ij\}}$ &$\bsymm $& $-1$ & $1$ \\
$G_{\mu}$ &{\rm Adj} & $0$ & $0$    \\
\end{tabular}
\end{minipage}
\begin{minipage}{2.2 in}
\begin{tabular}{c||ccc}
 & $SU(N)$ & $U_V(1)$ & $U_A(1)$  \\
  \hline \hline 
$\psi_{[ij]}$& $\asymm$ & $1$ & $1$  \\
 $\widetilde{\psi}^{[ij]}$ &$\basymm $& $-1$ & $1$ \\
$G_{\mu}$ &{\rm Adj} & $0$ & $0$    \\
\end{tabular}
\end{minipage}
\end{center}
\caption{The fermion sector of the orientifold  theories. $\psi$
and $\widetilde{\psi}$ are two Weyl fermions, while $G_{\mu}$
stands for the gauge bosons. In the left (right) parts of the
table the fermions are in the two-index symmetric (antisymmetric)
representation of the gauge group SU$(N)$. $U_V(1)$ is the
conserved global symmetry while the $U_A(1)$ symmetry is lost at
the quantum level due to the chiral anomaly.} \label{table}
\end{table}

The name of {\it orientifold field theory} is borrowed from
string-theory terminology. In fact, these theories were shown to
live on a brane configuration of type 0A string theory
\cite{Armoni:1999gc,Angelantonj:1999qg} which consists of NS5
branes, D4 branes and an {\it orientifold} plane. The gauge groups
in the parent and daughter theories are the same, and so are the
gauge couplings. 

In \cite{Sannino:2003xe} the effective Lagrangians for orientifold
theories were constructed in terms of the relevant low-lying color-singlet states.
The effective Lagrangians of this type have a long history
\cite{Schechter:1980ak,Rosenzweig:1979ay,Di Vecchia:1980ve,Witten:1980sp,Nath:1979ik,Salomone:1980sp,Kawarabayashi:1980dp,Migdal:1982jp,Veneziano:1982ah,Cornwall:1983zb,Cornwall:1984pa,Ellis:1984jv,Gomm:1984zq,Sannino:1999qe} and are known
to concisely encode nonperturbative aspects of strongly coupled
theories, such as the vacuum structure and symmetries, both exact
and anomalous \footnote{Among recent developments in this direction was the
demonstration of how the information on the center of the SU($N$)
gauge group (i.e.  $Z_N$) is efficiently transferred to the
hadronic states \cite{Sannino:2002wb}. This demonstration led to a
deeper understanding of the deconfining phase transition
\cite{Mocsy:2003tr} in pure Yang-Mills theory. When quarks were
added, either in the fundamental or in the adjoint representations
of the gauge group, a link between the chiral and deconfining
phase transitions was uncovered~\cite{Mocsy:2003qw}.}.

We will start by summarizing the low energy effective Lagrangians \cite{Sannino:2003xe} and then couple them non-minimally to gravity. Since the bosonic sector of the nonsupersymmetric orientifold field theories at large $N$ maps into the one of SYM we  identify first the gluino-ball state in SYM with the inflaton.  We then explore the consequences on the inflationary dynamics.  We will then include the $1/N$ corrections as well as the small gluino mass corrections. We investigate the inflationary parameters and check the consistency of our results against the slow-roll conditions and inflaton-inflaton scattering. We discover that the inflationary dynamics to the lowest order in $1/N$ is insensitive to corrections coming from the axial anomaly sector of the orientfiold field theories. However it does depend on the corrections to the vacuum energy and the guino mass term. Since the orientifold field theory with two-index antisymmetric matter for three colors is QCD with one flavor one can interpret the model as a new non-minimally coupled one-flavor QCD inflationary model. For all the models the compositeness scale is shown to be around $10^{16}$~GeV and the unitarity constraint from inflaton scattering is safely around the Planck scale.

\section{Nonminimal Super Yang-Mills Inflation}
\label{SYMI}
 
Before considering the coupling to gravity it is instructive to briefly review the construction of the SYM
effective Lagrangian while setting up the notation. 

\subsection{Super Yang-Mills Effective Action and Setup}
The Lagrangian of SU($N$) supersymmetric gluodynamics
is\,\footnote{The Grassmann integration is defined in such a way that
$\int \, \theta^2\, d^2\theta =2$.}
\be
{\cal L}& =&\frac{1}{4g^2}\,
\int \! {\rm d}^2 \theta\, \mbox{Tr}\,W^2 +{\rm H.c.} \nonumber\\[3mm]
& =&-\frac{1}{4 g^2} \,  G_{\mu\nu}^a G^{a\mu\nu}
+\frac{1}{2g^2}\, D^a D^a +\frac{i}{g^2} \lambda^a\sigma^\mu {\cal
D}_\mu\bar \lambda^a \, ,
\ee
where $g$ is the gauge coupling, the
vacuum angle is set to zero and 
\begin{equation} \mbox{Tr}\,W^2 \equiv
\frac{1}{2}W^{a,\alpha}W^{a}_{\alpha}= -
\frac{1}{2}\lambda^{a,\alpha}\lambda^{a}_{\alpha}\,. \end{equation}
The effective Lagrangian in supersymmetric gluodynamics was constructed by Veneziano and Yankielowicz (VY) \cite{Veneziano:1982ah}. In terms of the composite color-singlet chiral superfield $S$,
\begin{equation} S= \frac{3}{32\pi^2 N }\,\mbox{Tr}\,W^2 \,, \end{equation}
it reads
 \be {\cal L}_{VY}&=& \frac{9\, N^2}{4\,\alpha}\, \int d^2\!\theta\,
d^2\!{\bar{\theta}}
\left(S S^{\dagger}\right)^{\frac{1}{3}} 
+\frac{N}{3} \int \! {\rm d}^2\theta\, \left\{ S  \ln
\left(\frac{S }{\Lambda^3}\right)^{N}-NS\right\}  + \mbox{H.c.} \,
, \label{VY} \nonumber \\\ee
where $\Lambda$ is an invariant scale of the theory. The factor $N^2$ is singled out in the K\"{a}hler
term to make the parameter $\alpha$ scale as $N^0$, see
Eq.~(\ref{alphascaling}) below. The standard definition of the
fundamental scale parameter is \cite{Hagiwara:2002fs}
\beq \Lambda_{\rm st} =\mu \left(\frac{16\pi^2}{\beta_0 \,
g^2(\mu)} \right)^{\beta_1/\beta_0^2} \exp\left( -
\frac{8\pi^2}{\beta_0 \, g^2 (\mu)} \right)\ ,
\eeq
which for SYM theory is exact \cite{Novikov:1983uc}
and reduces to
\beq \Lambda^3_{\rm SUSY\,\,YM} =\mu^3\left(\frac{16\pi^2
}{3N\,g^2(\mu)}\right) \exp\left( -\frac{8\pi^2}{N \, g^2 (\mu)}
\right)\,.
\eeq
The exact value of the gluino condensate is due to the
 holomorphic property of SYM theory and in terms of
$\Lambda^3_{\rm SUSY\,\,YM}$ reads
\cite{Shifman:1999mv,Davies:1999uw}:
\beq
 \langle S\rangle = \frac{9}{32\pi^2}\Lambda^3_{\rm SUSY\,\,YM}\,.
 \eeq
Comparing with Eq.~(\ref{VY})  one deduces that
\beq \Lambda^3= \frac{9}{32\pi^2}\Lambda^3_{\rm SUSY\,\,YM}
\eeq
is $N$ independent. The gluino condensate
scales as $N$ as it should be. To determine the normalization of the constant $\alpha$ we
require the mass of the physical excitations to be $N$ independent,
\beq \alpha \sim {N^0} \,. \label{alphascaling}
\eeq
Indeed, the common mass of the bosonic and fermionic components of
$S$ is $M=2\alpha\, \Lambda /3$.  The chiral superfield $S$ at the
component level has the standard decomposition $S(y)=\varphi(y) +
\sqrt{2} \theta \Sigma(y) + \theta^2 F(y)$, where $y^\mu$ is the
chiral coordinate, $y^\mu=x^\mu - i \theta \sigma^\mu
\bar{\theta}$, and
\beq \varphi\ ,\quad \sqrt{2}\Sigma\ ,\quad F =\frac{3}{64\pi^2
N}\times \, \left\{
\begin{array}{l}
-\lambda^{a,\alpha}\lambda^{a}_{\alpha}\\[3mm]
G^a_{\alpha\beta}\lambda^{a,\beta} +2i D^a \lambda^{a}_{\alpha} \\[3mm]
-\frac{1}{2} G^a_{\mu\nu} G^{a\mu\nu}+
\frac{i}{2}G^a_{\mu\nu} \tilde{G}^{a\mu\nu}+\mbox{f.t.}
\end{array}
\right.
\label{decomp}
\eeq
where f.t. stands for (irrelevant) fermion terms.

The complex field $\varphi$ describes the scalar and pseudoscalar
gluino-balls while $\Sigma$ is their fermionic composite partner and the $F$ field is an auxiliary field.  To construct the low energy effective potential one uses the axial and trace anomalies. These are:
\begin{equation}
\partial^\mu J_\mu = \frac{N }{16\pi^2}\,
G_{\mu\nu}^a\tilde{G}^{a,\, \mu\nu}\, ,\qquad J_\mu= -
\frac{1}{g^2}\, \lambda^a \sigma_\mu \,\bar\lambda^a\, ,
\label{3anom}
\end{equation}
and
\beq \vartheta^\mu_{\mu}=- \frac{3 N }{32 \pi^2}\, G_{\mu\nu}^a
{G}^{a ,\, \mu\nu}\,\, ,
\eeq
where $J_\mu$ is the chiral current and $\vartheta^{\mu\nu}$ is
the standard  symmetric energy-momentum tensor.

In SYM theory these two anomalies belong to the same
supermultiplet \cite{Ferrara:1974pz} and hence, the coefficients
are the same (up to a trivial 3/2 factor due to normalizations).
In the orientifold theory the coefficients of the chiral and scale
anomalies coincide only at $N=\infty$; the subleading terms are
different.

Summarizing, the component bosonic form of the VY Lagrangian is: 
 \begin{equation}
\mathcal{L}_{\rm SYM}=\frac{N^{2}}{\alpha}\left(\varphi\overline{\varphi}\right)^{-\frac{2}{3}}g^{\mu\nu}\partial_{\mu}\varphi\partial_{\nu}\overline{\varphi}-V_{\rm SYM}\,,\,\,\,V_{\rm SYM}=\frac{4\alpha N^{2}}{9}\left(\varphi\overline{\varphi}\right)^{\frac{2}{3}}\ln\left(\frac{\varphi}{\Lambda^{3}}\right)\ln\left(\frac{\overline{\varphi}}{\Lambda^{3}}\right),\label{sym}
\end{equation}

with $\alpha$ the constant.  We consider this effective action to be the large $N$ limit of orientifold field theories and neglect all the fermionic degrees of freedom which are supposed to decouple in this limit. 

\subsection{Super Yang-Mills non-minimally coupled to gravity}

We are finally ready to take the next step and write the non-minimally coupled scalar component part of the superglueball action to gravity which in the Jordan frame reads:
\begin{align}
\mathcal{S}^{\rm J}_{\rm SYM}=\int d^{4}x\sqrt{-g}\left[-\frac{{\cal{M}}^{2}+N^2\xi\left(\varphi\overline{\varphi}\right)^{\frac{1}{3}}}{2}g^{\mu\nu}R_{\mu\nu}+\frac{N^{2}}{\alpha}\left(\varphi\overline{\varphi}\right)^{-\frac{2}{3}}g^{\mu\nu}\partial_{\mu}\varphi\partial_{\nu}\overline{\varphi}-V_{SGI}\right]\ . \label{sym1}
\end{align}

Before imposing the conformal transformation we concentrate on the modulus of $\varphi$ which we shall continue to call $\varphi$. In this case, we have
\begin{align}
\Omega^{2}=\frac{{\cal{M}}^{2}+N^2\xi\varphi^{\frac{2}{3}}}{M^{2}_{\rm P}}. \label{sym2}
\end{align}

The action in the Einstein frame reads:
\begin{align}
\mathcal{S}^{\rm E}_{\rm SYM}=\int d^{4}x\sqrt{-g}\left[-\frac{M^{2}_{\rm P}}{2}g^{\mu\nu}R_{\mu\nu}+\frac{N^{2}}{\alpha}\Omega^{-2}\left(1+ {\alpha f}\frac{N^2 \xi ^2}{3 M_{\rm P}^2} \Omega^{-2} \varphi^{\frac{2}{3}} \right)g^{\mu\nu} \varphi ^{-\frac{4}{3}}\partial_{\mu}\varphi\partial_{\nu}\varphi-\Omega^{-4}V_{\rm SYM}(\varphi)\right], \label{sym3}
\end{align}

where $f=1(0)$ is the metric (Palatini) formulation and  
\begin{align}
V_{\rm SYM}(\varphi)=\frac{4\alpha N^{2}}{9}\varphi^{\frac{4}{3}}\left(\ln\left(\frac{\varphi}{\Lambda^{3}}\right)\right)^{2} \ . \label{sym4}
\end{align}

We now introduce a canonically normalized field $\chi$ related to $\varphi$ via
\begin{align}
\frac{1}{2} \tilde{g}^{\mu \nu} \partial_{\mu} \chi (\varphi) \partial_{\nu} \chi(\varphi) = \frac{1}{2} \left( \frac{d \chi}{d \varphi} \right)^2 \tilde{g}^{\mu \nu} \partial_{\mu} \varphi \partial_{\nu} \varphi \ ,
\end{align}
with
\begin{align}
\frac{1}{2} \left( \frac{d \chi}{d \varphi} \right)^2 = \frac{N^{2}}{\alpha}\Omega^{-2}\left(1 +  {\alpha f}\frac{ N^2 \xi ^2}{3 M_{\rm P}^2} \Omega^{-2} \varphi^{\frac{2}{3}} \right) \varphi ^{-\frac{4}{3}}. \label{sym5}
\end{align}

In terms of the canonically normalized field we have: 
\begin{align}
\mathcal{S}^{\rm E}_{\rm SYM} &=\int d^{4}x \sqrt{-g}\left[-\frac{1}{2} M_{\rm P}^2 g^{\mu \nu}R_{\mu \nu} + \frac{1}{2} g^{\mu \nu} \partial_{\mu} \chi \partial_{\nu} \chi- U(\chi)  \right] \ , 
\end{align}
with
\begin{align}
U(\chi) \equiv \Omega^{-4}V_{\rm SYM}(\varphi) \ .
\end{align}
\subsection{Slow-roll parameters for non-minimally coupled super Yang-Mills}
We will analyze the dynamics in the Einstein frame, and therefore define the slow-roll parameters in terms of $U$ and $\chi$:
\begin{align}
\epsilon = \frac{M_{\rm P}^2}{2} \left( \frac{dU / d \chi}{U} \right)^2, \quad \quad \eta = M_{\rm P}^2 \left( \frac{d^2U / d \chi^2}{U} \right), \quad \quad {\cal N} = \frac{1}{M_{\rm P}^2} \int _{\chi_{end}} ^{\chi_{ini}} \frac{U}{dU /d\chi} d \chi\ . \label{sym6}
\end{align}
We consider here the large field regime, i.e.: 
 \begin{equation}
 \varphi^{\frac{2}{3}} \gg \frac{{\cal{M}}^2}{N^2\xi} \ . 
 \label{lfe}
 \end{equation}
In this limit $\epsilon$ becomes, in the $\varphi$ variable:
\begin{align}
\epsilon_{SYM} \simeq 
 \frac{1}{ \left(\ln \left( \frac{\varphi}{\Lambda ^3} \right)\right)^2 \left( \zeta^{-1} + f \cdot \frac{1}{3} \right)}\,,\quad\zeta^{-1}\equiv\frac{\xi^{-1}}{\alpha}\,.
 \label{zeta}
\end{align}

Inflation ends when $\epsilon_{SYM}=1$  such that:
\begin{align}
 \frac{\varphi_{\rm end}^{SYM}}{\Lambda^3} &= \exp \left( \frac{1}{ \sqrt{\left(\zeta^{-1}   +f \cdot \frac{1}{3}\right)}} \right) \label{sym7}.
\end{align}

In the large field limit the number of e-foldings  is:
\begin{align}
\mathcal{N} 
\simeq \frac{1}{2}\left[  \left(\zeta^{-1} +f \cdot \frac{1}{3} \right)\left( \ln \left( \frac{\varphi}{\Lambda^3}\right)\right)^2 \right]_{\varphi_{\rm end}}^{\varphi_{\rm ini}}.
\end{align}

A simple way to determine the value of $\varphi_{\rm ini}$ associated to when inflation starts is to require a minimal numbers of e-foldings compatible with a successful inflation, i.e. ${\cal N}=60$. This leads to: 
\begin{align} 
\frac{\varphi_{\rm ini}^{SYM}}{\Lambda^3} & \simeq \exp \left( \sqrt{\frac{121}{\zeta^{-1} +f \cdot \frac{1}{3}} }\right) \ .
\end{align}

Further relevant information can be extracted using the  WMAP \cite{arXiv:0812.3622} normalization condition:
\begin{align}
\frac{U_{\rm ini}}{\epsilon^{\rm SYM}_{\rm ini}} = (0.0276\,M_{\rm P})^4.
\end{align}

The label {\it ini} signifies that this expression has to be evaluated at the beginning of the inflationary period. This condition helps estimating the magnitude of the non-minimal coupling. We deduce: 
\begin{align}
U_{\rm ini}  
 \simeq \frac{4\alpha\,M_{\rm P}^4}{9  N^{2} \xi^2} \left(\ln \left( \frac{\varphi_{\rm ini}^{\rm SYM}}{\Lambda^3}\right)\right)^{2}  
\simeq \frac{4\alpha\,M_{\rm P}^4}{9  N^{2} \xi^2}\left(\frac{121}{\zeta^{-1} +f \cdot \frac{1}{3}}\right)\ .
\end{align} 
and
\begin{align}
\epsilon^{\rm SYM}_{\rm ini} \simeq  \frac{1}{ \left(\ln \left( \frac{\varphi_{\rm ini}^{\rm SYM}}{\Lambda ^3} \right)\right)^2 \left(\zeta^{-1}  + f \cdot \frac{1}{3} \right)}\simeq\frac{1}{\left(\frac{121}{\zeta^{-1} +f \cdot \frac{1}{3}}\right) \left( \zeta^{-1} + f \cdot \frac{1}{3} \right)}= 0.0083\ .
\end{align}
We can therefore determine the magnitude of the non-minimal coupling which depends, in principle, on whether we use the Palatini ($f=0$) or the metric formulation ($f=1$). In the case of the Palatini formulation we have:  
\begin{align}
N^2 \xi  \simeq 1.1 \times 10^{10}\alpha^{2} \equiv \xi_P \quad\quad \text{Palatini}  
\end{align}
 { The situation for the metric case turns out to be subtle because of the interplay between the structure of the non-minimal coupling to gravity and the large $N$ counting.  In the limit in which $\zeta^{-1}\equiv\frac{\xi^{-1} }{\alpha}\ll  \frac{1}{3}$ we find:}  
\begin{align}
N \xi \simeq 1.83 \times 10^{5} \sqrt{\alpha} = \xi_m \quad \quad \text{Metric}\quad{\rm with}\,\quad\,\zeta^{-1}\equiv\frac{\xi^{-1}}{\alpha} \ll  \frac{1}{3}.
\label{Nfixed}
\end{align}
{With $\alpha$ of order unity we can still allow for relatively large values of $N$ satisfying \eqref{Nfixed}. 
The phenomenologically large value of $\xi$ is common to the case of Higgs inflation \cite{Bezrukov:2007ep}, and other earlier approaches \cite{Spokoiny:1984bd,Futamase:1987ua,Salopek:1988qh,Fakir:1990eg,Komatsu:1999mt,Tsujikawa:2004my} . A more complete treatment for all these models would require to discover in the future a mechanism for generating such a large coupling.} 

The knowledge of the non-minimal coupling allows us to estimate the initial and final value of the composite glueball field $\varphi$ which reads: 
\begin{align}
&\frac{{\varphi_{\rm end}^{\rm SYM}}^{\frac{1}{3}}}{\Lambda} \sim e^{\frac{\sqrt{\alpha\xi_P}}{3N}}, \qquad \qquad  \frac{{ \varphi_{\rm ini}^{\rm SYM}}^{\frac{1}{3}}}{\Lambda} \sim e^{\frac{\sqrt{121\alpha\xi_P}}{3N}} \qquad \text{Palatini} 
&& \nonumber \\
&\frac{{\varphi_{\rm end}^{\rm SYM}}^{\frac{1}{3}}}{\Lambda} \sim 1.8 , \qquad \qquad \,\,\,\,\,\, \frac{{\varphi_{\rm ini}^{\rm SYM}}^{\frac{1}{3}}}{\Lambda}\sim 570 \qquad ~~~~~~~\text{Metric}\quad{\rm with}\,\quad\,\zeta^{-1}\equiv\frac{\xi^{-1}}{\alpha} \ll  \frac{1}{3}.\end{align}
In the large $N$ limit  (say $N\simeq { \sqrt{\alpha \xi_P}}/{30}$)  also the Palatini formulation leads to initial and final values for $\varphi$ within a few times $\Lambda$. 

It is possible to further relate the strongly coupled scale $\Lambda$ with the Planck mass recalling that  in the large field regime \eqref{lfe} we expect on/near the ground state $N^2 \xi \Lambda^2 \simeq M^2_{\rm P}$. Assuming for the reduced Planck mass the value $2.44\times 10^{18}$ GeV we obtain
\begin{equation}
\Lambda  \simeq \frac{0.57}{\sqrt{N}\alpha^{1/4}} \times 10^{16}~ 
\text{GeV} \ .
\end{equation} 
These results are encouraging and indicate that it is possible to conceive an inflationary scenario driven by a SYM-like composite inflaton. This value is not only consistent with the results found in \cite{Channuie:2011rq,Bezrukov:2011mv} but shows that it is possible to lower the scale of composite inflation by increasing the number of underlying colors. We recall that $\alpha$ is  given by the underlying theory and is expected to be of order unity \cite{Feo:2004mr}. 

\subsection{Inflaton scattering and its unitarity constraint}
\label{unitarity}

Next, we turn to the interesting question of the constraints set by tree-level unitarity of the inflaton field. According to the potential given above, the ground state reads:
\begin{equation}
 \langle \varphi \rangle = \Lambda^3 \equiv v^3 \ . \label{SGI-14}
\end{equation}

It is worth noting here that the potential evaluated on the ground state has zero energy. In addition, we are interested in the large field regime which can be well approximated by setting $\mathcal{M}=0$. At the minimum of the potential, the following relation naturally holds: 
\begin{equation}
 M^{2}_{\rm P} \simeq N^2 \,\xi v^{2} \,\,\,\, \Rightarrow \,\,\,\, \Omega = \frac{\varphi^{\frac{1}{3}}}{v} \ . \label{SGI-15}
\end{equation}

For later convenience in this section, we introduce the field $\phi$ possessing unity canonical dimension related to $\varphi$ as follows
\begin{equation}
\varphi=\phi^{3} \ . \label{SGI-16}
\end{equation}

In the Einstein frame, we obtain
\begin{align}
\mathcal{S}^{\rm E}_{\rm SYM}=\int d^{4}x\sqrt{-g}\left[-\frac{M^{2}_{\rm P}}{2}g^{\mu\nu}R_{\mu\nu}+\frac{9 N^{2}}{2\alpha}\frac{v^{2}}{\phi^{2}}\left(2+ \frac{2}{3} f\zeta \right)g^{\mu\nu}\partial_{\mu}\phi\partial_{\nu}\phi-4 N^{2}\alpha \, v^{4}\left[\ln\left(\frac{\phi}{v}\right)\right]^{2}\right]. \label{SGI-17}
\end{align}
Violation of tree-level unitarity of the scattering amplitude concerns the inflaton field fluctuations  $\delta \phi$ around its classical time dependent background $\phi_c(t)$ during the inflationary period
\begin{equation}
 \phi(\vec{x},t) = \phi_c(t) + \delta \phi(\vec{x},t) \ .
 \end{equation}
In first approximation it is possible to neglect the time dependence of the classical field and write 
\begin{equation}
 \phi(\vec{x}\,) = \phi_c + \delta \phi(\vec{x}\,) \ .
 \end{equation}
To estimate the actual cutoff of the tree-level scattering amplitude we analyze independently the kinetic and potential term for the inflaton in the Einstein frame. Expanding the kinetic term around the classical background we obtain: \begin{equation}
 \frac{3N^{2}}{2\alpha}\frac{v^2}{\phi_c^2}\left(6+2 f\zeta\right) (\partial \delta \phi)^2 
 \sum_{n=0}^{\infty} (n+1) \frac{(-\delta\phi)^n}{\phi_c^n} \ .
 \label{Kint}
 \end{equation}
It is possible to canonically normalize the first term of the series, i.e. the kinetic term for a free field, by rescaling the fluctuations as follows:
\begin{equation}
\frac{ \delta \phi}{ \phi_c} = \frac {\sqrt{\alpha} \,\,\delta \widetilde{\phi} }{\sqrt{3} N v\sqrt{\left(6+2 f\zeta \right)}}\ . 
\end{equation} 

Under this field redefinition we find:
\begin{equation}
\frac{1}{2} (\partial \delta \widetilde{\phi})^2\sum_{n=0}^{\infty} (n+1) \frac{(-\sqrt{\alpha}\,\,\delta \widetilde{\phi})^n}{ (18 + 6 f \zeta)^{\frac{n}{2}}{(N\,v)}^n} \ .
 \label{normKint}
\end{equation}

For the potential term  the higher order operators are also, respectively, of the form: 
\begin{equation}
\frac{(\sqrt{\alpha}\,\,\delta \widetilde{\phi})^{n}}{ (18 + 6 f \zeta)^{\frac{n}{2}}(N\,v)^{n}}\ ,
\end{equation}
%

This implies that the tree-level cutoff for unitarity is: 
\begin{equation}
\sqrt{\frac{18 + 6  f \zeta}{\alpha}} \, N\, v \ .\end{equation}
This results shows that the cutoff is background independent. In the metric formulation the cutoff  is $N v \sqrt{6\xi} \sim \sqrt{6} M_{\rm P} $, i.e. a little higher than the Planck scale. This implies that the theory is valid, from the unitarity point of view, till the Planck scale. 

\section{Orientifold Inflation}
\label{OI}
We wish now to deform the supersymmetric effective action to describe inflation driven by the gauge dynamics of SU($N$) gauge theories with one Dirac fermion in either the two-index antisymmetric or symmetric representation of the gauge group. Following \cite{Sannino:2003xe} we start by recalling  the trace and axial anomalies for the orientifold theories: 
\begin{eqnarray}
\vartheta^{\mu}_{\mu} &=&2N\left[N \pm
\frac{4}{9}\right]\left(F + \bar F\right) = -3\left[N \pm
\frac{4}{9}\right]\frac{1}{32\pi^2}\, G_{\mu\nu}^a {G}^{a ,\, \mu\nu} \ , \label{trace}\\[3mm]
\partial^{\mu} J_{\mu}
&=&i\,\frac{4N}{3}\,\left[ N \mp 2\right]\, \left(\bar F - F\right)=
\left[N \mp 2\right]\frac{1}{16\pi^2}\, G_{\mu\nu}^a {\tilde{G}}^{a
,\, \mu\nu} \ , \label{axial}
\end{eqnarray}
where the top (bottom) sign is for the antysimmetric (symmetric) theory and 
\begin{eqnarray}
\varphi= -\frac{3}{32\pi^2\, N}\,
 \widetilde{\psi}^{\alpha,[i,j]_{\mp}} \psi_{\alpha,[i,j]_{\mp}} \ ,
\label{phi1}
\end{eqnarray}
and $F$ is given in  Eq.~(\ref{decomp}).  The gluino field of
supersymmetric gluodynamics is replaced here by  two Weyl fields, $ \widetilde{\psi}^{\alpha,[i,j]_{\mp}}$ and
$\psi_{\alpha,[i,j]_{\mp}}$, which can be combined into one Dirac
spinor.  The top (bottom) sign for the bracket in $ \widetilde{\psi}^{\alpha,[i,j]_{\mp}} $ indicates antisymmetric (symmetric) color indices. The color-singlet field $\varphi$ is bilinear in $
\widetilde{\psi}^{\alpha,[i,j]_{\mp}}$ and  $\psi_{\alpha,[i,j]_{\mp}}$. 
\subsection{$1/N$ - Orientifold Effective Action and then Inflation}
\label{tre-finite-N}
In this limit we can drop subleading $1/N$ terms in the
expressions for the trace and chiral anomaly. Then it is clear that  the anomalous currents map into
the ones of SYM. Therefore the effective action built to saturate at the tree level trace and axial anomaly 
has the same form as in Eq. (\ref{sym}).  Hence, by keeping only the leading-$N$ terms only one recovers the supersymmetry-based bosonic properties, i.e. degeneracy of the opposite-parity mesons and the vanishing of the vacuum energy. Of course in this limit the symmetric and the antisymmetric orientifold theories are  indistinguishable.

To parametrize the $1/N$ corrections at the effective Lagrangian level we use the results of \cite{Sannino:2003xe} and write: 

\beq
 {\cal L}_{\rm OI}={\cal F}(N)\left\{
\frac{1}{\alpha}\left(\varphi\, \overline{\varphi} \right)^{-2/3}\,
\partial_\mu\overline{\varphi}\,\partial^\mu\varphi-\frac{4\alpha}{9}\,
\left(\varphi\, \overline{\varphi} \right)^{2/3}\,\left(
\ln\overline{\Phi}\,\ln\Phi - \beta \right)\right\}\,,
\label{fnocomponent}
\eeq
where $\beta $ is a numerical real and positive \cite{Sannino:2003xe}
parameter,
\beq
\beta = O(1/N)\,, \eeq and \beq {\cal F}(N) = N^2(1 + \beta^{\prime})
\qquad  {\rm with}  \qquad \beta^{\prime} ={\cal O}(1/N) \ .
\eeq
However the sign of $\beta^{\prime}$ is not known. In \cite{Sannino:2003xe} one did not have to take into account the leading $1/N$ corrections to ${\cal F}(N)$ since this function drops out from any physical quantity. However when coupled to gravity these corrections cannot be neglected.
We have also: 
\beq \Phi =
\varphi^{1+\epsilon_1}\,\overline{\varphi}^{-\epsilon_2}\,,\qquad
\overline{\Phi}= \overline{\varphi}^{1+\epsilon_1}\,\varphi^{-\epsilon_2}\,,
\label{newfields} \eeq where $\epsilon_{1,2}$ are parameters of order
$O(1/N)$, \beq \epsilon_1 = \mp \frac{7}{9\,N}\,,\qquad \epsilon_2
= \mp \frac{11}{9\,N}\,. \label{newparam}
\eeq
The top (bottom) sign corresponds to the two-index antisymmetric (symmetric) theory. The scale and chiral dimensions of $\overline{\Phi}$ and $\Phi$ are engineered to saturate the axial and trace anomalies  for the orientifold theories.

For the purpose of investigating the inflationary paradigm we restrict the potential to the real part of the field $\varphi$ and write: 
\begin{align}
\mathcal{L}_{\rm OI} \rightarrow  \frac{{\cal F}(N)}{\alpha}\left(\varphi\right)^{-4/3}g^{\mu\nu}
\partial_{\mu}{\varphi}\partial_{\nu}{\varphi} - V_{OI}\left( {\varphi} \right) \ ,
\end{align}
with 
\begin{align}
V_{\rm OI} \left( \varphi \right)= \frac{4 \alpha'}{9} \varphi^{\frac{4}{3}}  \left( \ln \left(\frac{\varphi}{\Lambda^3}\right) ^2 - \frac{\beta}{ \left(1 + \epsilon_A \right)^2} \right), \quad \alpha' \equiv {\cal F}(N) \alpha \left(1 + \epsilon_A \right)^2,\quad \epsilon_A \equiv \epsilon_1 - \epsilon_2 \ . \label{OI-Pot}
\end{align}
To leading order in $1/N$ we have: 
\begin{equation}
\alpha^{\prime} = N^2 \, \alpha \left( 1 + 2 \epsilon_A + \beta^{\prime}+ {\cal O}(1/N^2) \right) \ , \qquad \frac{\beta}{(1+\epsilon_A^2)} = \beta + {\cal O}(1/N^2) \ .
\label{alphap}
\end{equation}
Adding gravity we have: 
\begin{align}
\mathcal{S}^{\rm J}_{\rm OI}=\int d^{4}x\sqrt{-g}\left[-\frac{{\cal{M}}^{2}+ N^2 \xi \varphi^{\frac{2}{3}}}{2}g^{\mu\nu}R_{\mu\nu}+{\cal L}_{OI}\right]\ . \label{LOI}
\end{align}
For the non-minimal coupling to gravity we have assumed, for simplicity, the same used for SYM. Neglected terms in $1/N$ in the non-minimal coupling could be re-absorbed in a redefinition of the function ${\cal F}$. With this choice the only $1/N$ corrections come from the gauge sector. 

The action in the Einstein frame reads:
\begin{align}
\mathcal{S}^{\rm E}_{\rm OI}=\int d^{4}x\sqrt{-g}\left[-\frac{M^{2}_{\rm P}}{2}g^{\mu\nu}R_{\mu\nu}+\frac{{\cal F}(N)}{\alpha}\Omega^{-2}\left(1+ \frac{\alpha f N^4}{{\cal F}(N)}\frac{ \xi ^2}{3 M_{\rm P}^2} \Omega^{-2} \varphi^{\frac{2}{3}} \right)g^{\mu\nu} \varphi ^{-\frac{4}{3}}\partial_{\mu}\varphi\partial_{\nu}\varphi-\Omega^{-4}V_{\rm OI}(\varphi)\right] \ , \label{sym3}
\end{align}
with the same $\Omega$ as in the SYM case. 
\subsection{Orientifold slow low parameters}
In the large field regime $\varphi^{2/3} \gg \frac{{\cal{M}}^{2}}{N^2\xi}$ we obtain:
\begin{align}
\epsilon_{OI} \simeq \epsilon_{SYM}\left[
1 + \frac{2}{\left({\ln \frac{\varphi}{\Lambda^3}}\right)^2} \beta - \frac{3}{3 + f   \,\alpha \xi}\,\beta^{\prime}  \right]\ , \quad {\rm with} \quad \epsilon_{SYM} = \frac{1}{\left(\ln \frac{\varphi}{\Lambda^3}\right)^2 \left( \zeta^{-1} + \frac{f}{3}\right)} \ ,
\end{align}
and  $\zeta^{-1} = \xi^{-1}/\alpha$ defined first in  \eqref{zeta}. 
The value of the field at the end of inflation can be determined by setting $\epsilon_{OI}(\varphi_{end}) =1$. 
We use perturbation theory in the small parameters $\beta$ and $\beta^{\prime}$ and search for a solution to this condition of the type: 
\begin{equation}
\varphi_{end} = \varphi_{end}^{SYM} + \beta \varphi_1 + \beta^{\prime} \varphi_{1}^{\prime}\ , \quad {\rm with} \quad \varphi_{end}^{SYM} = \Lambda^3 e^{\frac{\sqrt{3\alpha\xi}}{\sqrt{3 + f\alpha\xi }}}\ .
\end{equation}
The solution reads: 
\begin{eqnarray}
\varphi_{end} = \varphi^{SYM}_{end}\left[1 -   \frac{1}{2\zeta\left(\zeta^{-1} + \frac{f}{3}\right)^{3/2}}  \beta^{\prime} + 
 \sqrt{\zeta^{-1} + \frac{f}{3}} \,\beta
\right] \ .
\end{eqnarray}
 The number of e-foldings reads: 
\begin{align}
{\cal N}_{OI} & \simeq  
 \left[ \frac{\left(\ln \frac{\varphi}{\Lambda^3}\right)^2}{2\zeta} \left( 1 - 
\frac{2 \ln\ln \frac{\varphi}{\Lambda^3}}{\left(\ln\frac{\varphi}{\Lambda^3}\right)^2}\beta +\beta^{\prime}\right) \right]_{\varphi_{\rm end}}^{\varphi_{\rm ini}} \ .
\end{align}

We fix the initial value of the inflaton field $\varphi_{\rm ini}$ by requiring a total of $60 $ e-foldings during inflation and obtain: 
\begin{eqnarray}
\varphi_{ini} = \varphi^{SYM}_{ini}\left[1+ (1 + \ln(11)) \sqrt{\zeta^{-1} + \frac{f}{3}}  \, \frac{\beta}{11}  -  
 \frac{11}{2\zeta\left(\zeta^{-1} + \frac{f}{3}\right)^{3/2}} \,\beta^{\prime}
\right] \ .
\label{iniO}
\end{eqnarray}
For $\zeta^{-1} \ll 1/3$ and in the metric case we get the following range for inflation:
\begin{align}
\frac{\varphi_{ ini}^{1/3}}{\Lambda} \simeq 570\,(1+ 0.06  \, \beta) , \quad \quad \frac{\varphi_{end}^{1/3}}{\Lambda} \simeq 1.8 \,(1 + 0.19 \, \beta)  \quad \quad \text{Metric}.
\end{align}
To take the limit above we have assumed $\zeta$ large at any $N$ although strictly speaking at extremely large $N$ this approximation may break down. However for any large but finite $N$ we expect this result to hold.


Perhaps the most relevant result is that the slow-roll parameters are insensitive to the breaking of holomorphicity induced by $\epsilon_A$ i.e. the corrections coming from the mismatch between trace and axial anomaly coefficients. The irrelevance of $\epsilon_A$ is due to the fact that all the slow-roll parameters are defined as ratios of derivatives of the potential divided by the potential itself. And $\epsilon_A$ appears only in a function multiplying the overall potential to leading order in $1/N$. Therefore the results are valid for both orientifold field theories.

\section{  One-flavor QCD Inflation}
\label{OFI}
Another way to depart from a supersymmetric theory is to add soft supersymmetric breaking operators such as the mass for the gluino. Following Masiero and Veneziano \cite{Masiero:1984ss} one can therefore add the gluino mass term 
\begin{equation}
\Delta {\cal L}_m = - \frac{m}{2g^2}\, \lambda^{\alpha} \lambda_{\alpha} + {\rm h.c.} \ .
\end{equation}
At the effective Lagrangian level, it reads
\begin{equation}
\Delta {\cal L}_m = 4 \frac{m}{2 \lambda } N^2 \left(\varphi + \overline{\varphi} \right) 
\end{equation}
with $\lambda \equiv g^2 N/8\pi^2$ the 't Hooft coupling. We assume here that the mass parameter is real and positive. If this were not the case one can render it real and positive by redefining the vacuum angle $\theta$. We will also assume that softness condition $m/\lambda \ll \Lambda$. One can, however, start immediately from the orientifold theory where the mass term reads 
\begin{equation}
\Delta {\cal L}_m = - \frac{m}{g^2}\, \psi^{\alpha} \widetilde{\psi}_{\alpha} + {\rm h.c.} \ .
\end{equation}
The color indices for the gluino and the orientifold field theories are (implicitly) contracted to obtain color singlet operators, while the spin indices are explicit and contracted. In the large $N$ limit this term, at the effective Lagrangian level, maps exactly in the one above \cite{Sannino:2003xe}. 
Since for three colors the orientifold field theory with antisymmetric matter is QCD with one flavor we can therefore study non-minimally coupled inflation driven by a QCD-like theory even featuring a light fermion mass.  

The effective Lagrangian augmented with the quark mass reads  \cite{Sannino:2003xe}
\begin{align}
\mathcal{L}_{\rm 1F-QCD}={\cal F}(N)\left[\frac{1}{\alpha}\left(\varphi\bar{\varphi}\right)^{-2/3}g^{\mu\nu}\partial_{\mu}\bar{\varphi}\partial_{\nu}\varphi-\frac{4\alpha}{9}\left(\varphi\bar{\varphi}\right)^{2/3}\left(\ln\bar{\Phi}\ln\Phi-\beta\right)\right]+\frac{4m}{3\lambda}N^{2}\left(\varphi+\bar{\varphi}\right)\,, \label{N-m}
\end{align}

where
\begin{align}
\Phi=\varphi^{1+\epsilon_{1}}\bar{\varphi}^{-\epsilon_{2}},\,\,\bar{\Phi}=\bar{\varphi}^{1+\epsilon_{1}}\varphi^{-\epsilon_{2}},\,\,\epsilon_{1}=\frac{7}{9N},\,\,\epsilon_{2}=\frac{11}{2N},\,\,\beta=\mathcal{O}(1/N)\,. \label{SGI-19}
\end{align}

\begin{align}
V_{\rm 1F-QCD}= \frac{4 \alpha'}{9} \varphi^{\frac{4}{3}}  \left( \ln \left(\frac{\varphi}{\Lambda^3}\right) ^2 -\beta \right) - \frac{8 N^2 m'}{3 } \varphi, \qquad  {\rm with } \qquad m^{\prime} =  \frac{m}{\lambda} .
\end{align}
Here $\alpha^{\prime}$ assumes the same form of \eqref{alphap}. The associated slow-roll parameter epsilon expanded at the leading order in $\beta$, $\beta'$, $\epsilon_A$ and $m^{\prime}$ reads: 
 \begin{equation}
 \epsilon_{1F-QCD} \simeq \epsilon_{SYM}\left[
1 + \frac{2}{\left({\ln \frac{\varphi}{\Lambda^3}}\right)^2} \beta - \frac{3}{3 + f   \,\alpha \xi}\,\beta^{\prime}     + \frac{2 \left(6+ \ln \left( \frac{\varphi}{\Lambda^3}\right) \right)}{\varphi^{1/3} \alpha  \left( \ln \left( \frac{\varphi}{\Lambda^3}\right)\right)^2}  m^{\prime} \right]\ ,   
 \end{equation}
 The dependence on $\epsilon_A \equiv \epsilon_1 - \epsilon_2$ is subleading to the order we are investigating and therefore does not appear here. This is so since it would necessarily come multiplied by $m^{\prime}$. 
 
Imposing that the end of inflation occurs for $\epsilon_{1F-QCD} = 1$ we obtain to the first order in $\beta$, $\beta^{\prime}$ and $m^{\prime}$
\begin{eqnarray}
\varphi_{end} = \varphi^{SYM}_{end}\left[1 -   \frac{1}{2\zeta\left(\zeta^{-1} + \frac{f}{3}\right)^{3/2}}  \beta^{\prime} + 
 \sqrt{\zeta^{-1} + \frac{f}{3}} \,\beta +
  \frac{\alpha^{-1}}{\exp\left[{\left( {3\sqrt{\xi^{-1} + \frac{f}{3}}}\right)^{-1}} \right]} \left( 1 + 6 \sqrt{\xi^{-1} + \frac{f}{3}}\right) m^{\prime}
\right] \ . \nonumber \\
\end{eqnarray}
In the metric formulation it collapses to: 
\begin{eqnarray}
\varphi_{end} = \Lambda^3 \, e^{\sqrt{3}}\left[1+ 
 \frac{\beta}{\sqrt{3}} +    
  \frac{\left( 1 + 2\sqrt{3} \right)}{\alpha \, e^{\frac{1}{\sqrt{3}}}}  m^{\prime}
\right] \ ,  \qquad {\text{Metric}} \nonumber \\
\end{eqnarray}
anticipating that $\xi$ assumes a very large value. 

However, for the initial value of the inflaton, obtained by requiring 60 efoldings, we get the same results for the coefficients of $\beta$ and $\beta^{\prime}$ as in \eqref{iniO} while the coefficient for $m^{\prime}$ is cumbersome to write down explicitly.  We therefore opt for providing the result  directly in the metric formulation which reads 
\begin{eqnarray}
\varphi_{ini} \simeq \varphi^{SYM}_{ini}\left[1+ \frac{1 + \ln(11)}{11\sqrt{3}} \, \beta + \frac{0.46}{\alpha} m^{\prime}\right] \ .
\end{eqnarray}

In the metric approach we have the following range for inflation: 
\begin{align}
\frac{\varphi_{ ini}^{1/3}}{\Lambda} \simeq 570\,(1+ 0.06  \, \beta + \frac{0.15}{\alpha} m^{\prime}) \  , \quad \quad \frac{\varphi_{end}^{1/3}}{\Lambda} \simeq 1.8 \,(1 + 0.19 \, \beta + \frac{0.83}{\alpha} m^{\prime})  \quad \quad \text{Metric}.
\end{align}
These results show that the inflationary slow-roll parameters are not sensitive to the axial anomaly departure from the supersymmetric limit. Furthermore we learn that the effects of a small nonzero negative vacuum energy induced by the presence of a positive and real $\beta$ term, of order $1/N$, leads to higher values of the inflaton field with respect to the SYM values. Finally the effects of a quark mass are similar to the $1/N$ corrections, albeit the sign of the corrections are sensitive to the $\theta$ angle choice. Here we have chosen a value of $\theta$ leading to the same sign of the $\beta$ term.
 \section{Conclusions}
We explored the paradigm of non-minimally coupled composite inflation further by considering orientifold field theories. We have shown that the advantage of using these theories is that at large number of colors they share certain super Yang-Mills properties. Because of these properties we were able to use for inflation the bosonic part of the Veneziano-Yankielowicz effective theory. We have include the $1/N$  and fermion mass corrections at the effective Lagrangian level. This allowed us to determine the associated corrections on the inflationary slow-roll parameters. Additionally we showed that the scale of composite orientifold inflation is of the order of $10^{16}$~GeV and that unitarity for inflaton scattering leads to a cutoff at the Planck scale.

\end{document}